\documentclass[aps,showpacs,pre,twocolumn,reprint]{revtex4-1}
\usepackage{graphicx,bm,amsmath,amssymb,natbib,url}
\usepackage{dcolumn}
\usepackage[colorlinks=true,linkcolor=blue,allcolors=blue]{hyperref}

\usepackage{color}
\usepackage[normalem]{ulem}
\definecolor{darkgreen}{rgb}{0.1,.6,.1}

\allowdisplaybreaks

\begin{document}

\title{Effects of time-varying habitat connectivity on metacommunity persistence}

\author{Subhendu Bhandary${}^{1}$}
\author{Debabrata Biswas${}^{2}$}
\author{Tanmoy Banerjee${}^{3}$}\email{tbanerjee@phys.buruniv.ac.in}
\author{Partha Sharathi Dutta${}^{1}$}
\thanks{Corresponding author}
\email{parthasharathi@iitrpr.ac.in}

\affiliation{\vspace{0.1in} ${}^{1}$Department of Mathematics, Indian Institute of
  Technology Ropar, Rupnagar 140001, Punjab,
  India\\  ${}^{2}$Department of Physics, Bankura University, Bankura 722155, West Bengal, India\\  ${}^{3}$Chaos and Complex Systems Research
  Laboratory, Department of Physics, University of Burdwan, Burdwan
  713104, West Bengal, India\\ }%

\received{:to be included by reviewer}
\date{\today}

\begin{abstract}

Network structure or connectivity pattern is critical in determining collective dynamics among interacting species in ecosystems. Conventional research on species persistence in spatial populations has focused on static network structure, though most real network structures change in time, forming time-varying networks. This raises the question, in metacommunities, how does the pattern of synchrony vary with temporal evolution in the network structure. The synchronous dynamics among species are known to reduce metacommunity persistence. Here, we consider a time-varying metacommunity small-world network consisting of a chaotic three-species food chain oscillator in each patch/node. The rate of change in the network connectivity is determined by the natural frequency or its subharmonics of the constituent oscillator to allow sufficient time for the evolution of species in between successive rewirings. We find that over a range of coupling strengths and rewiring periods, even higher rewiring probabilities drive a network from asynchrony towards synchrony. Moreover, in networks with a small rewiring period, an increase in average degree (more connected networks) pushes the asynchronous dynamics to synchrony. On the contrary, in networks with a low average degree, a higher rewiring period drives the synchronous dynamics to asynchrony resulting in increased species persistence. Our results also follow the calculation of synchronization time and robust across other ecosystem models. Overall, our study opens the possibility of developing temporal connectivity strategies to increase species persistence in ecological networks.  

\end{abstract}

\maketitle 

\section{Introduction}

Synchronization among populations of the same species is a widely observed collective phenomenon in the studies of ecological networks \cite{blasius1999complex,HoHa08}. Synchrony in the dynamics of populations creates interdependence in their abundance, and simultaneous low abundance can lead to simultaneous extinction. Thus, synchrony increases the risk of network-wise extinction and correspondingly reduces species persistence. Population synchrony can be driven by several factors, including dispersal network structures or connectivity patterns \cite{HoHa08,gupta2017increased}. Even though species connectivity via dispersal has attracted much attention due to its both positive and negative effects on the persistence and stability of spatially separated populations \citep{koelle2005dispersal,gravel2011persistence,fox2017population,dutta2015spatial}, how temporal changes in the connectivity can influence species persistence have received much less attention. Specifically, how temporal changes in species connectivity influences nonlinear dynamics of a metacommunity is still unclear \citep{pilosof2017multilayer}. \citet{leibold2004metacommunity} has defined a {\em metacommunity} as ``a set of local communities that are linked by dispersal of multiple potentially interacting species".

Connectivity between spatially separated habitat patches is an integral component of metacommunity ecology \citep{walther2002ecological,HoHa08,hodgson2009climate,senior2019global}. \citet{taylor1993connectivity} described `connectivity' among habitat patches as ``$\dots$connectivity is the degree to which the landscape facilitates or impedes movement among resource patches''. Over time, many approaches have led to alternative definitions of population connectivity, e.g.,~structural, genetic, and functional connectivity \citep{kool2013population}. Though population connectivity can be defined in various ways under diverse ecological circumstances, they share a common characteristic that corresponds to spatial linkages/dependencies between populations or individuals. Many studies have shown that population connectivity via dispersal is as important to population viability as distribution of resources \citep{fahrig1988effect}; however, connectivity patterns in fragmented landscapes are in general ignored. Landscapes, where species movements occur, can vary temporally through distribution and quality of habitat over time \citep{zeigler2014transient}. As a result, species dynamics can vary along the complex spectrum of `static' to `dynamic' environments \citep{levins1969some,Han99}. Metacommunity dynamics in static environments have mostly focused on static networks, where links offer a permanent connectivity pattern between habitat patches \citep{moilanen1998long,HoHa08}. However, habitats are disturbance-driven in dynamic landscapes, and links between them are best described to form `temporal' networks, where connectivity may change across different timescale \citep{bishop2018evaluating}. For example, the marsh fritillary butterfly {\em Euphydryas aurinia} in Finland, inhabiting dynamic landscapes, exhibits patch networks that vary over time \citep{wahlberg2002dynamic}.

In a temporal patch network, the links/connectance between species habitat patches varies over time \citep{holme2012temporal,pilosof2017multilayer,sundaresan2007network}. More specifically, a temporal patch network is a `sequence of separate networks' on the same set of patches/nodes, where each such snapshot is characterized by an adjacency matrix (i.e., a square matrix representing the structure of a finite graph/network) for a particular time duration \citep{li2017fundamental}.   Therefore, temporal connectivity can also be recognized with a `transient' feature. For dynamic environments, \citet{zeigler2014transient} describe that structure of connectivity should be seen as time-varying (transient) rather than static, due to changes in biotic and abiotic conditions influencing metacommunity dynamics. Temporal connectivity pattern is also known to create a short window during which temporal opportunities for movement between particular patches increase depending on a species' generation time or life history. Nonetheless, to understand metacommunity dynamics governing by the changes in the species interaction patterns due to their life history or anthropogenic factors, temporal networks could provide a useful framework \citep{olesen2008temporal,mucha2010community,olesen2011strong}.

For static network structures, it is known that an increased dispersal strength inevitably induces a higher degree of synchrony and ultimately reduces metapopulation persistence \citep{hastings1993complex,blasius1999complex}. \citet{koelle2005dispersal} have shown that by adopting a metacommunity framework, this pattern of persistence can be altered, resulting in dispersal induced de-synchronization. Exploring the nature of synchronization/de-synchronization dynamics under the framework of temporal networks has led to exciting observations in generic networks of nonlinear units \cite{Boccaletti06,Sorrentino08,Ghosh22}. The basin stability measure \cite{menck2013basin} has been used to determine the stability of the synchronous state in temporal networks \cite{kohar2014}. \citet{masuda2013temporal} describes that synchronization is more challenging to achieve in temporal networks than in the corresponding aggregate networks. Surprisingly, most of the studies deal with networks whose structure changes faster than the characteristic timescale of their individual units. In this fast-changing network structure, the dynamics of the system may be considered as static in terms of synchronization stability under the adiabatic approximation \cite{Stilwell06,Porfiri06,kohar2014,Petit17}. However, in ecological networks, change in network structure occurs at a much slower rate, which can be best predicted by the dominant period or the corresponding harmonics of oscillations of individual nodes.

Motivated by the above arguments, in this paper, we study the synchronous/asynchronous dynamics of an ecological time-varying network whose time rate of change (or the rewiring frequency) is comparable with the natural frequency or its subharmonics of the constituent nodes. We consider the small-world network topology \cite{strogatz2001} as the core network structure, and the uncoupled dynamics of the nodes are governed by the chaotic Hastings-Powell model of the three-species food chain \cite{hastings1991chaos}. 
Here, we employ the wavelet transform method to identify each node's dominant period of oscillation and its harmonics,  and the network is rewired following those periods. Moreover, appropriate coupling strengths are chosen based on the master stability function approach \cite{pecora1998master}. Importantly, for suitable coupling strength, average degree, and rewiring period, we find that an increase in the rewiring probability drives the network from asynchronous to synchronous state; however, further increase of rewiring probability eventually leads to asynchronous dynamics. We also find that temporal networks with a higher average degree and small rewiring period can propel the asynchronous dynamics to a synchronous one and, therefore, reduce species persistence. Our results are supported by measures from master stability function \cite{pecora1998master} and the basin stability \cite{menck2013basin}. We further corroborate our results using the concept of clustering frequency and the transient time of synchronization.  Finally, we demonstrate the generality of our study through another temporal network, where the species dynamics in each node are governed by the  Blasius-Huppert-Stone foodweb model \cite{blasius1999complex}.

\section{Models and Methods \label{s:model}} 

\subsection{A metacommunity model}

We study the dynamics of a metacommunity model consisting of $N$ spatially separated patches connected by dispersal that follows a time-varying network topology. In each patch, the uncoupled dynamics are governed by a chaotic three-species food chain model \citep{hastings1991chaos}; with a basal resource population ($x$), an intermediate consumer population ($y$), and a top predator population ($z$). Within the patch, dynamics of the food chain are characterized by the logistic growth function and the type-II functional response. Further, diffusive dispersal connects the interacting patches, which forms a time-varying network described by the following set of differential equations:
\begin{subequations}\label{eq1}
  \begin{align} 
\frac{dx_i}{dt} & = x_i(1-x_i)-\frac{a_1 x_i y_i}{1+b_1 x_i},\\ 
\frac{dy_i}{dt} & = \frac{a_1 x_i y_i}{1+b_1 x_i}- \frac{a_2 y_i
  z_i}{1+b_2 y_i}-d_1 y_i +\epsilon_1 \sum_{j=1}^{N} L_{ij}
y_j,\\ 
\frac{dz_i}{dt} & = \frac{a_2 y_i z_i}{1+b_2 y_i}-d_2 z_i+\epsilon_2 \sum_{j=1}^{N} L_{ij} z_j,
\end{align} 
\end{subequations}
where $i (=1,2,...,N)$ describes the node/patch index. Here, the consumer $y$ depends on the resource $x$ for its survival, and the predator $z$ at the top level depends on the consumer $y$. The system parameters of the uncoupled model (i.e., when $\epsilon_1=0$ and $\epsilon_2=0$) are: $a_1$, $a_2$, $b_1$, $b_2$, $d_1$ and $d_2$. Unless stated, throughout this paper we consider the parameter values of the uncoupled model as: $a_1 = 5$, $a_2 = 0.1$, $b_1 = 3$, $b_2 = 2$, $d_1 = 0.4$, and $d_2 = 0.01$ \citep{hastings1991chaos}.

The diffusive dispersal connects the patches with dispersal rates $\epsilon_1$ and $\epsilon_2$ for the consumer ($x$) and the top predator ($y$), respectively. For simplicity, in this study we have assumed $\epsilon_1 = \epsilon_2 = \epsilon$. Here, both species immigration and emigration are described by the Laplacian matrix ($L_{ij}$) obtained from the adjacency matrix ($A_{ij}$) of the considered network. In particular, elements of the adjacency matrix are defined as: $A_{ij}=1$, if patches $i$ and $j$ are connected via dispersal; and otherwise $A_{ij}=0$. The diagonal elements of the Laplacian matrix are the sum of columns (or rows) of the adjacency matrix with the negative sign, representing the emigration from the $i$-th patch to other connected patches. In other words, the diagonal elements of the Laplacian matrix is the degree of each $i$-th patch with the negative sign, i.e., $L_{ii} = -\sum_{j=1}^{N} L_{ij}$~(for each $i)~= -$ the degree of $i$-th node, and $L_{ij}=A_{ij}$ when $i \neq j$.

\subsection{Temporal-network with each snapshot following a small-world network topology}

Various network structures can be considered depending on the connectivity pattern between spatially separated patches, such as regular, small-world, and random networks, in the metacommunity model (\ref{eq1}). These network structures are widely used in ecology, and other fields to study the collective dynamics of coupled oscillators \cite{ranta2007population,HoHa08,stankovski2017coupling,ArDu18,arumugam2019dynamic}. Each of these network structures can be generated by the Watts-Strogatz algorithm \citep{WaSt98} for different values of a rewiring probability ($p$). For example, a network is regular if $p=0$, completely random when $p=1$, and follows a small-world structure if $0<p<1$.

\begin{figure}[!ht]
\centering
\includegraphics[width=0.98\columnwidth,angle=0]{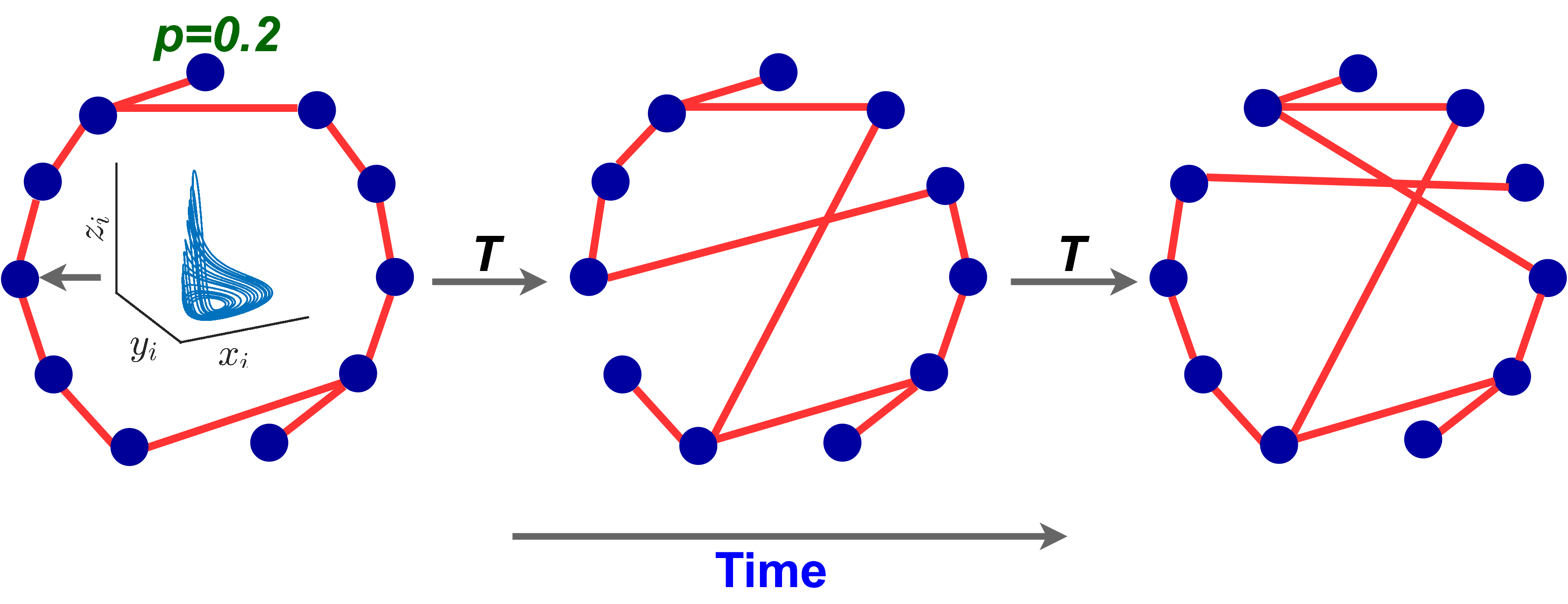}
\caption{Schematic representation of a time-varying network composed with a `sequence of separate networks'. Each sub-figure represents a snapshot that follows a small-world network topology associated with a rewiring probability ($p$). After a fixed period (say $T$), there is a change in the network structure, keeping the rewiring probability unaltered. A chaotic dynamical system governs the uncoupled dynamics in each node.} \label{f:schme}
\end{figure}

Traditional research on ecological networks has considered small-world and random network structures under the framework of static networks \cite{ranta2007population,HoHa08}. In a static network, the connectivity structure is invariant over time. However, in a temporal network, the connectivity evolves involving two key mechanisms, i.e., when and how the connectivity changes. Here, we study the collective dynamics of the metacommunity model (\ref{eq1}) that follows a temporal network structure and is composed of a chaotic oscillator at each patch. For the sake of completeness and comparison, we also study the system's dynamics for static network structure. Figure~\ref{f:schme} demonstrates a schematic representation of our modeling framework. The initial network is chosen after rewiring a regular network with the probability ($p$). The patch connectivity is rewired at each fixed period ($T$), keeping the rewiring probability unaltered. Although we allow connectivity to evolve between patches at a fixed time interval $T$, the average degree in the network remains unaltered. Hence, for a time-varying network, the Laplacian matrix ($L_{ij}$) in model \eqref{eq1} intermittently varies at each period $T$; otherwise, in the intermediate time, it remains unaltered. The following section discusses the choice of the rewiring period ($T$).
\begin{figure*}
\centering 
\includegraphics[width=0.95\textwidth,angle=0]{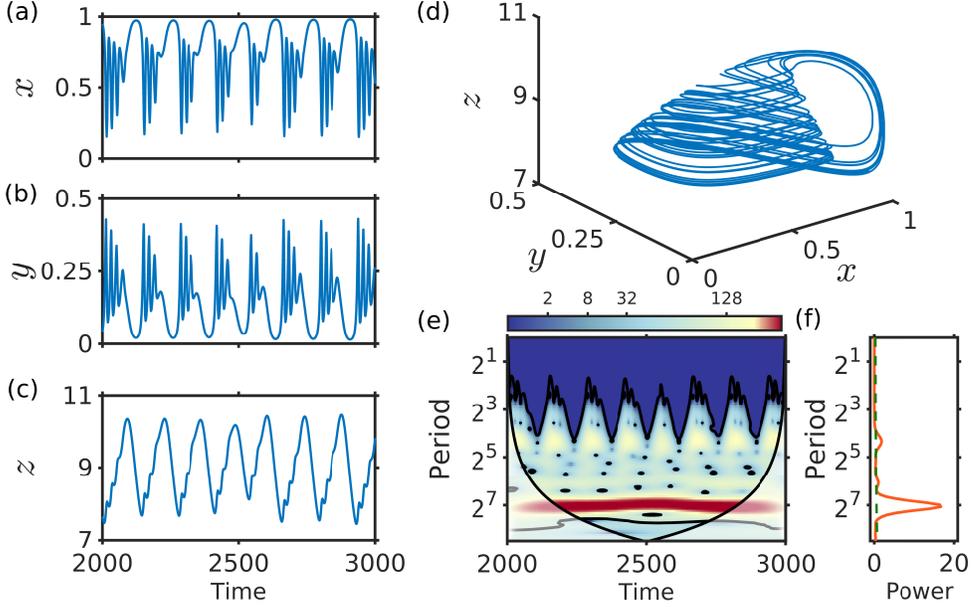}  
\caption{Wavelet analysis to a chaotic time series of the model \eqref{eq1} in the absence of coupling. Chaotic time series of (a) the resource, (b) the consumer, and (c) the top predator; corresponding (d) phase portrait of the chaotic attractor, (e) wavelet power spectra, and (f) wavelet global spectrum. Model parameters are $a_1=5$, $a_2=0.1$, $b_1=3$, $b_2=2$, $d_1=0.4$, and $d_2=0.01$.}
\label{f:time_phase}
\end{figure*}

\subsection{Wavelet analysis of a chaotic time series}

Unlike most studies on temporal networks, here we do not consider changes in the network structure at every integration step size. We rewire the network structure at a rate with a period $T$, which is determined by the characteristic time of the nodal oscillators. To determine $T$, we employ the wavelet transform to a chaotic time series of the uncoupled Hastings-Powell model. The chaotic time series consisting of multi-cycles (see Figs.~\ref{f:time_phase}(a)-\ref{f:time_phase}(c)) is analyzed through wavelet transform, which determines the localized variations within time series \citep{torrence1998}. From Figs.~\ref{f:time_phase}(e) and \ref{f:time_phase}(f), the dominant period and associated subharmonics of the chaotic time series can be found. From a practical point of view, the dominant period is comparable to the life cycle of a species; keeping this in mind, we rewire the networks at a subharmonic of the dominant period assuming that a species may change its dispersal networks structure a few times in a life cycle. Later in this paper, we show the importance of choosing the rewiring period ($T$).

\subsection{Linear stability analysis of synchronized solutions \label{M:MSF}}

The interaction/coupling strength plays a crucial role in governing the collective dynamics of a system of coupled oscillators. It is known that in the weak coupling regime, decreasing the coupling strength may weaken synchrony. Therefore, it is essential to know the suitable coupling range where the synchronous solution is stable. To determine the appropriate coupling range for the model \eqref{eq1}, here we follow the master stability function (MSF) approach \cite{pecora1998master}. Below, we briefly describe the MSF approach for temporal networks.

Consider a coupled system of identical oscillators written as: $\dot{X_{i}}=F(X_{i})$, $i=1,2,\dots,N$, $X_{i}\in R^{d}\rightarrow R^{d}$, $F:R^{d}\rightarrow R^{d}$, where $X_{i}$ represents the $d$-dimensional vector which describes the dynamics at the $i$-th node. At each isolated node of the network, the dynamics are governed by the function $F(X_{i})$. If each node interacts with its neighbours, then the dynamics of the $i$-th node can be written as: 
\begin{eqnarray}
\dot{X_{i}} &= & F(X_{i})+\epsilon \sum_{j=1}^{j=N} A_{ij}(t)[H(X_{j})-H(X_{i})],\nonumber \\
& = & F(X_{i})+\epsilon \sum_{j=1}^{j=N} L_{ij}(t)H(X_{j});\; i=1,..,N, \label{Eq:2C}
\end{eqnarray}
where $\epsilon$ represents the coupling strength, $L_{ij}$ is the Laplacian matrix, and $H: R^{d}\rightarrow R^{d}$ defines the coupling function representing the interaction between different nodes. Further, we calculate the local asymptotic stability of the oscillators along the synchronization manifold $X_{1}=X_{2}=X_{3}=\dots=X_{N}=X_{0}$. The variational equation of \eqref{Eq:2C} is given by:
\begin{equation}\label{eq:3P}
\dot{\xi}=[I_{N}\otimes DF+\epsilon L(t)\otimes DH]\xi,
\end{equation}
where $\xi=(X_{1}-X_{0}, X_{2}-X_{0}, X_{3}-X_{0}, \dots, X_{N}-X_{0})^{T}$ is the perturbation vector, $I_{N}$ is the $N \times N$ identity matrix, $\otimes$ represents the Kronecker product, $DF$ and $DH$ are the Jacobian function of $F$ and $H$, respectively, evaluated on the synchronous solution ($X_{0}$). If the Laplacian matrices $L(t)$ and $L(t')$ commute for any $t$ and $t'$, then we can find an orthogonal matrix $Q$ such that $Q^{T}L(t)Q$ is diagonal for all $t$, where $Q^T$ stands for transpose of $Q$. Using the block diagonalization form of \eqref{eq:3P} we obtain $N$ independent $d$ dimensional equation:
\begin{equation}\label{eq:4P}
\dot{\delta_{i}}=[DF+\epsilon  \lambda_{i}(t) DH] \delta_{i}, ~~i=1,\dots,N,
\end{equation}
where $(\delta_{1},\delta_{2},\dots, \delta_{N})^{T}=(Q^{T}\otimes I_{d})\xi$, and $\lambda_{i}$ are eigenvalues of $L$. The synchronous solution is stable, if all perturbation modes transverse to the synchronization manifold decaying asymptotically to zero. Decoupled variation equations \eqref{eq:4P} differ in $\lambda_{i}(t)$ and others terms are equal. To study the stability of the synchronous state it is enough to study the maximum Lyapunov exponent of \eqref{eq:4P} which is a function of $\alpha$.  
\begin{equation}
\dot{\zeta}=[DF+\alpha DH] \zeta.
\end{equation}
Here, $\alpha$ is the function of the eigenvalues $\lambda_{i}$ and coupling strength $\epsilon$, also known as the MSF and denoted by $\Lambda(\alpha)$. The synchronous solution is stable if the MSF $\Lambda(\alpha)$ is negative for all transverse modes ($i\geq 2$). Further, there are mainly three cases possible for $\Lambda(\alpha)<0$: (i) no such $\alpha$ exists: $\Lambda(\alpha)$ has no crossing point (ii) $\alpha_{1}<\epsilon \lambda_{i}$: $\Lambda(\alpha)$ has one crossing point, (iii) $\alpha_{1}<\epsilon \lambda_{i}<\alpha_{2}$: $\Lambda(\alpha)$ has two crossing points \citep{huang2009generic}.

Structural evolution in complex temporal networks has been studied more often via different rewiring techniques, such as slow switching (rewiring links after longer periods) and fast switching (more frequent rewiring). The condition for a stable synchronous state varies for slow and fast switching. Let a network switches among $M$ different configurations (snapshots) ${L_{1}, L_{2},\dots, L_{M}}$ after certain rewiring time period $T$, then the necessary condition for achieving stable sync state is \cite{zhou2016synchronization}:
\begin{align*}
\sum_{k=1}^{M} \dfrac{1}{M}\Lambda(\epsilon \lambda_{k}^{i_{k}})<0.
\end{align*}
If the network switch at a fast scale yielding $M$ arbitrary sequential structures, then the condition of stable sync state is as follows:
\begin{align*}
\Lambda(\dfrac{1}{M}\sum_{k=1}^{M} \epsilon \lambda_{k}^{i_{k}})<0.
\end{align*}
For a fast switching instance, stability of the synchronous state in a network with time-varying topology can be obtained by calculating the MSF for the static time-averaged network. Hence, when network structure evolves via fast switching, calculating the MSF from the time average of matrix $\bar{L}=\dfrac{1}{M}\sum_{k=1}^{M} L_{k}$ is sufficient \citep{stilwell2006sufficient}. Thus the type of switching scheme favorable for synchronization can be anticipated from the MSF approach pertaining to the switching variants. Thereafter, a concave(convex) MSF shape indicates that the network supports synchronization dynamics under a fast (slow) switching \cite{zhou2016synchronization}.

\subsection{Basin stability}

The basin stability (BS) is a non-local and nonlinear measure of stability related to the basin volume of multistable systems, including higher-dimensional complex networks \cite{menck2013basin}. The BS measure is known to complement the linear stability analysis. To determine the BS of the considered system \eqref{eq1}, we numerically simulate it for different initial conditions ($I$), chosen uniformly from the region $[0,1]\times[0,0.5]\times[7.5,11.5]$ (which has been chosen from the existence region of the chaotic attractor depicted in Fig.~\ref{f:time_phase}). If $I_{s}$ is the number of initial conditions that arrives at the synchronous state, then we define the $\mbox{BS} = \dfrac{I_{s}}{I}$. Whether an initial condition is converging to a synchronous state or not has been determined by an order parameter, namely the synchrony measure ($\sigma_m$) evaluated for a large enough time $\hat t$. The synchrony measure ($\sigma_m$) is defined as below \citep{KoMu11,ArDu18}:
\begin{eqnarray*}\label{syncmeaseq}
\sigma_m &= &\sqrt{1 - \left \langle \frac{\sum_{i=1}^{N}[X_i(t)-\overline{X(t)}~]^2}{\sum_{i=1}^{N}{X_i(t)}^2} \right \rangle},
\end{eqnarray*}
where $\overline{X(t)}=\frac{1}{N} \sum_{i=1}^{N}{X_i(t)}$, and $ \langle \dots \rangle$ denotes the average over the time period $\hat t$.  The synchrony measure $\sigma_m$ varies between $0$ and $1$. In particular, $\sigma_m=1$ denotes complete synchronization (perfect synchrony), $\sigma_m=0$ denotes no synchrony, and $0<\sigma_m<1$ marks partial synchrony.

The BS can change depending on the coupling strength and structural properties of a network. For each set of parameters, using $10^4$ initial conditions, we compute the BS in static and time-varying networks. In each case, after removing the transients, we use the measure $\sigma_m$ to identify whether the metacommunity is synchronized or not.

\subsection{Cluster identification}

The cluster analysis \cite{HoHa08,gupta2017increased} is used to study the coherence dynamics between a pair of patches $(i,j)$ of the metacommunity model \eqref{eq1}. Specifically, we calculate the linear correlation coefficient ($\rho_{ij}$) to compare the dynamics between a pair of patches $(i,j)$. Here, by considering the top predator populations ($z$) from patches $i$ and $j$, the pairwise linear correlation coefficient ($\rho_{ij}$) is computed as:
\begin{eqnarray*}
\rho_{ij} &=& \frac{\langle z_i z_j\rangle - \langle z_i\rangle \langle z_j\rangle}{\sqrt{\langle z_i^2\rangle - \langle z_i\rangle^2} \sqrt{\langle z_j^2\rangle - \langle z_j\rangle^2}},
\end{eqnarray*}
where $\langle \dots \rangle$ denotes the average over the time interval $[t,~t+{\hat t}]$, $\hat t$ denotes a long enough fixed time-period.  The $i$-th and $j$-th patches form a cluster whenever $\rho_{ij} \approx 1$. By calculating $\rho_{ij}$ for all pairs of patches, the number of clusters in each simulation of the time-varying network (with $N$-nodes) can be identified. Here, 1-cluster denotes global (perfect) synchrony, whereas $N$-cluster denotes complete asynchrony. Also, the time-varying network might exhibit ${n}$-clusters, where $1\leq n \leq N$.  Using these, we compute the frequency of the $n$-cluster, where the frequency at time $t$ is defined as:
\[\mbox{Frequency of}~ n\mbox{-cluster solution} =
 \frac{\mbox{No. of} \leq n\mbox{-clusters}}{\mbox{No. of simulations}}\;.\]

This will be useful, in particular, to understand the intermediate solutions (no of clusters between 2 to $N-1$); other than complete synchrony and asynchrony. The degree of metacommunity persistence can be understood from the cluster identification. Note that the linear correlation coefficient $\rho_{ij}$ is different from the synchrony measure $\sigma_m$, in the sense that the synchrony measure characterizes the coherent behavior among all the interacting patches, whereas the correlation coefficient characterizes the coherent behavior between two patches ($i$ and $j$). 
\begin{figure*}
\centering
\includegraphics[width=0.85\textwidth,angle=0]{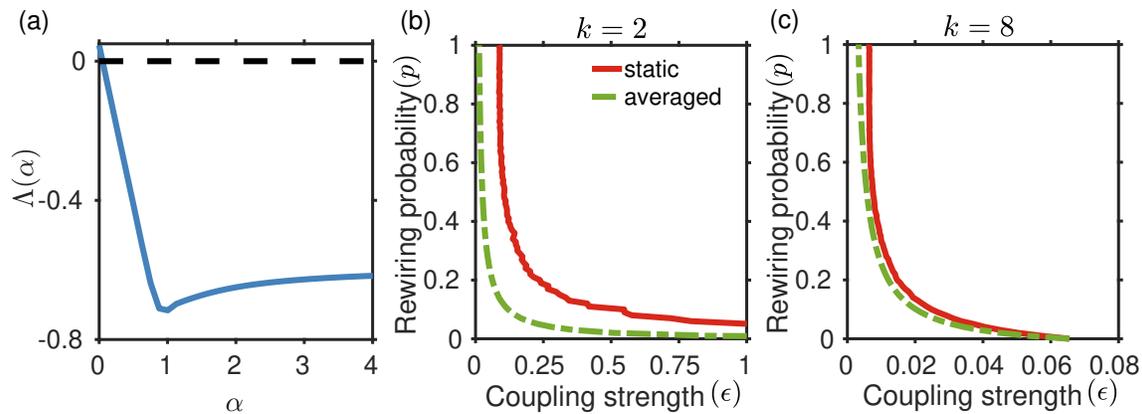}
\caption{(a) Master stability function for the metapopulation model \eqref{eq1}. The black dashed line marks the neutral line. Using the MSF approach, the coupling range of stable synchronized solution is calculated with variations in the rewiring probability for both the static and averaged networks; in (b) $k = 2$ and (c) $k = 8$. The region bounded below by the dashed (solid) curve marks the region of stable synchronous state for the averaged (static) network. As observed, on increasing the average degree ($k$), stable synchronous state is achieved even for lower coupling strength.}
\label{f:smt1}
\end{figure*}

\subsection{Synchronization time}

For a fixed rewiring period ($T$), the rewiring probability ($p$) and the coupling strength ($\epsilon$) simultaneously affect the coherence dynamics of a temporal network. The effect of variable $p$ and $\epsilon$ on the occurrence of complete synchrony can be determined by calculating the synchronization time \cite{kohar2014}. The time to reach the synchronous state in a complex network is known as the synchronization time ($S_t$). Indeed, the synchronization time divides the network dynamics into transient and asymptotic states. In the transient state (i.e., $t \in [0,~S_t]$), the dynamics of a time-varying network fluctuate between synchronous and asynchronous states, whereas in the asymptotic case, only synchronized dynamics exists (i.e., $t>S_t$). Hence, we compute the synchronization time whenever the network shows complete synchrony. While calculating the $S_t$, to determine network synchrony, we have used the synchrony measure ($\sigma_m$) for each small sub-intervals of a time series. In particular, when successive values of $\sigma_m$ in sub-intervals reach the maximum value ($\sigma_m=1$), the time of the first sub-intervals is denoted as the synchronization time of that particular network.

\section{Results}

\subsection{Determining the coupling range of stable synchronous solution using the MSF approach}

We start our analysis by calculating the coupling range in which the synchronous solution of model \eqref{eq1} is stable according to the MSF approach (discussed in Subsection~\ref{M:MSF}). From the MSF depicted in Fig.~\ref{f:smt1}(a), we find that the temporal network can stably synchronize below a critical coupling strength after crossing the zero line. For different values of the average degree $k$, Figs.~\ref{f:smt1}(b)-\ref{f:smt1}(c) illustrate the coupling range in which the synchronous solution for static as well as averaged networks are stable, with variation in the rewiring probability. We see that, for the averaged network, the range of stable synchronous state is broader than that of the static network. This also holds good for other metacommunity models (see Fig.~\ref{f:smt3} in the Appendix). Hence, the temporal network outperforms the static network in terms of synchronization stability. Further, with an increase in the average degree, there is an increase in the coupling range, i.e., the minimum coupling strength at which the synchronized state is stable decreases further with an increase in the average degree. The difference between the coupling ranges of static and averaged networks minimizes when the average degree increases (see Fig.~\ref{f:smt1}(c)).

\subsection{Synchronous and asynchronous dynamics in static and time-varying networks following the BS measure}

\begin{figure*}
\centering
\includegraphics[width=0.38\textwidth,angle=0]{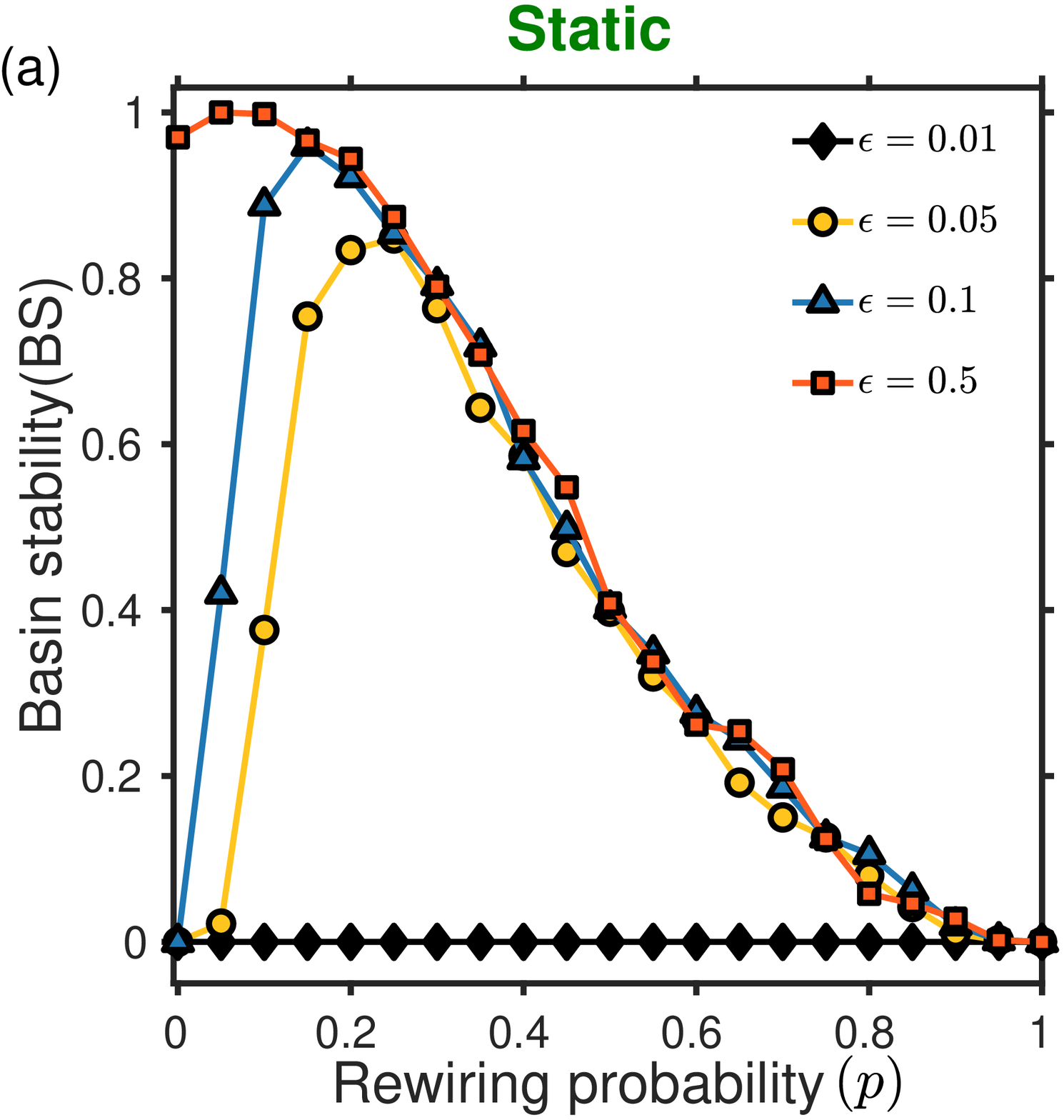} \hspace{0.2in}
\includegraphics[width=0.38\textwidth,angle=0]{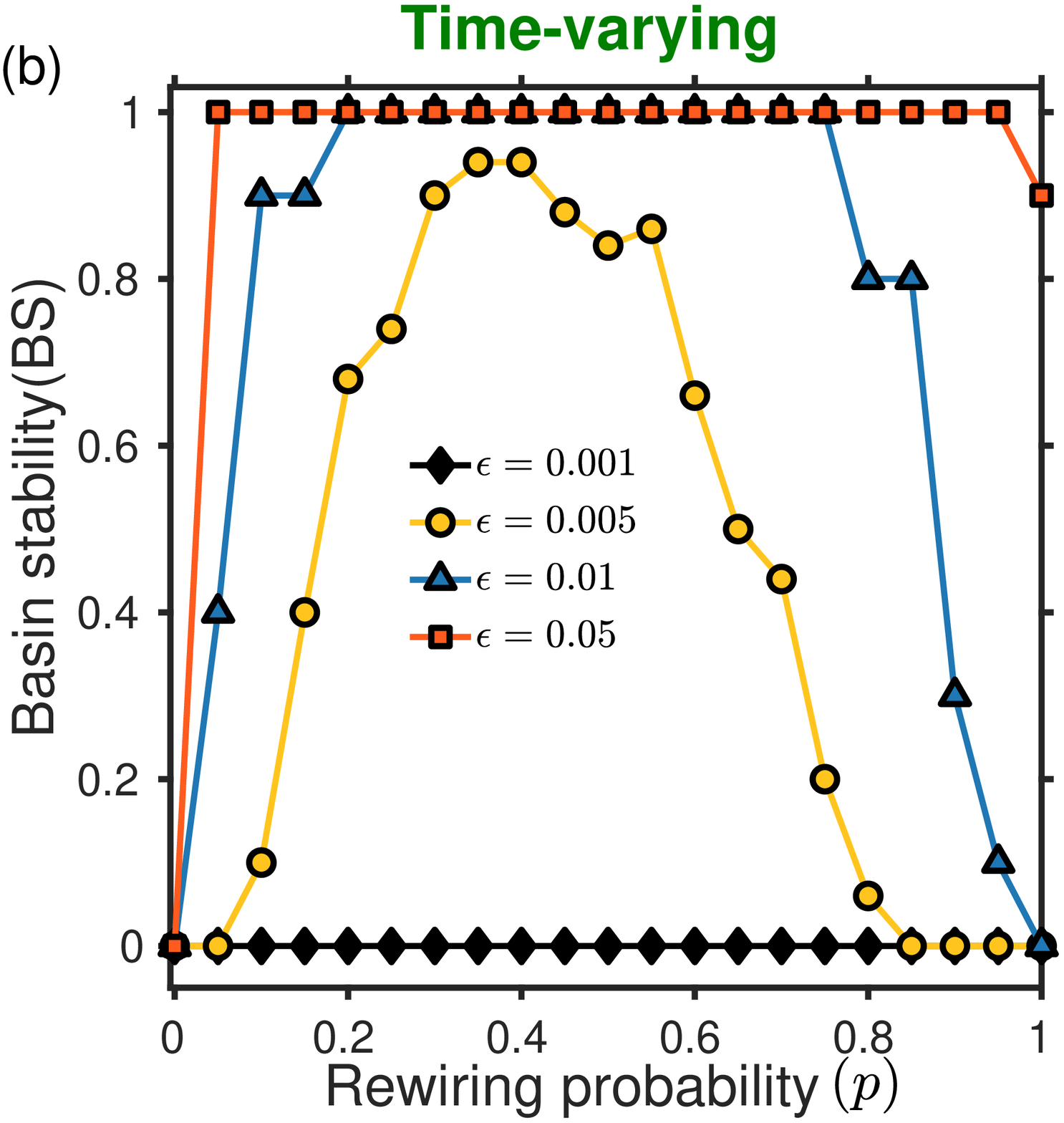}
\caption{Basin stability of (a) static, and (b) time-varying networks with variations in the rewiring probability ($p$), for different coupling strengths ($\epsilon$). At each value of $p$, the basin stability is computed using $10^4$ independent simulations in the time-interval $[0,10^4]$. Other parameters are $N=100$, $k=2$, and $T=16$.} \label{f:fig3}
\end{figure*}
\begin{figure*}
\centering
\includegraphics[width=0.925\textwidth]{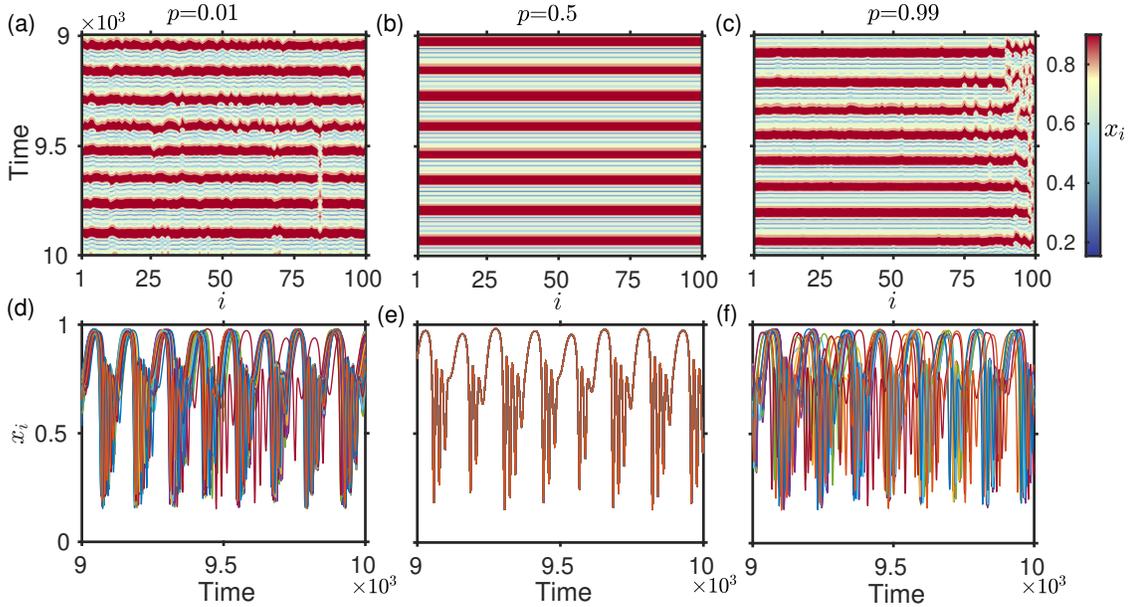}
\caption{Spatiotemporal dynamics and corresponding time series of the metacommunity model \eqref{eq1} with a temporal network structure: (a),(d) asynchronous oscillations for $p=0.01$; (b),(e) synchronous oscillations for $p=0.5$; and (c),(f) asynchronous oscillations for $p=0.99$. The other parameter values are: $N=100$, $k=2$, $\epsilon=0.005$, and $T=16$.}
\label{f:hp_tv_ts}
\end{figure*}

To understand the influence of network structure on collective dynamics of the metacommunity model \eqref{eq1}, we start with studying static networks. We consider static networks (i.e., $L_{ij}$ remains unchanged with time) that follow the Watts-Strogatz (WS) network topology with a rewiring probability ($p$). Then we consider static networks with $p$ value ranging from $p=0$ (regular) to $p=1$ (completely random). For each $p$ value, $10^3$ networks are generated, and the corresponding synchronous dynamics are analyzed in a time interval $[0,10^4]$ for an $\epsilon$. We calculate the BS measure to analyze synchrony in the metacommunity (see Fig.~\ref{f:fig3}(a)). We find that, as the $p$ value increases, for moderate values of $\epsilon$, the BS first increases and eventually decreases to zero (see Fig.~\ref{f:fig3}(a)). Therefore, for moderate values of $\epsilon$, random networks yield lower synchronization regions than a regular network, increasing the metacommunity persistence. However, as expected for weak $\epsilon$ values, the BS remains at zero, and there is no synchronization region. Our results are in agreement with previous literature - increasing randomness in a static network structure through the rewiring probability $p$ decreases the metacommunity synchronization and hence increases species persistence \citep{ranta2007population}.
\begin{figure*}
\centering
\includegraphics[width=0.38\textwidth]{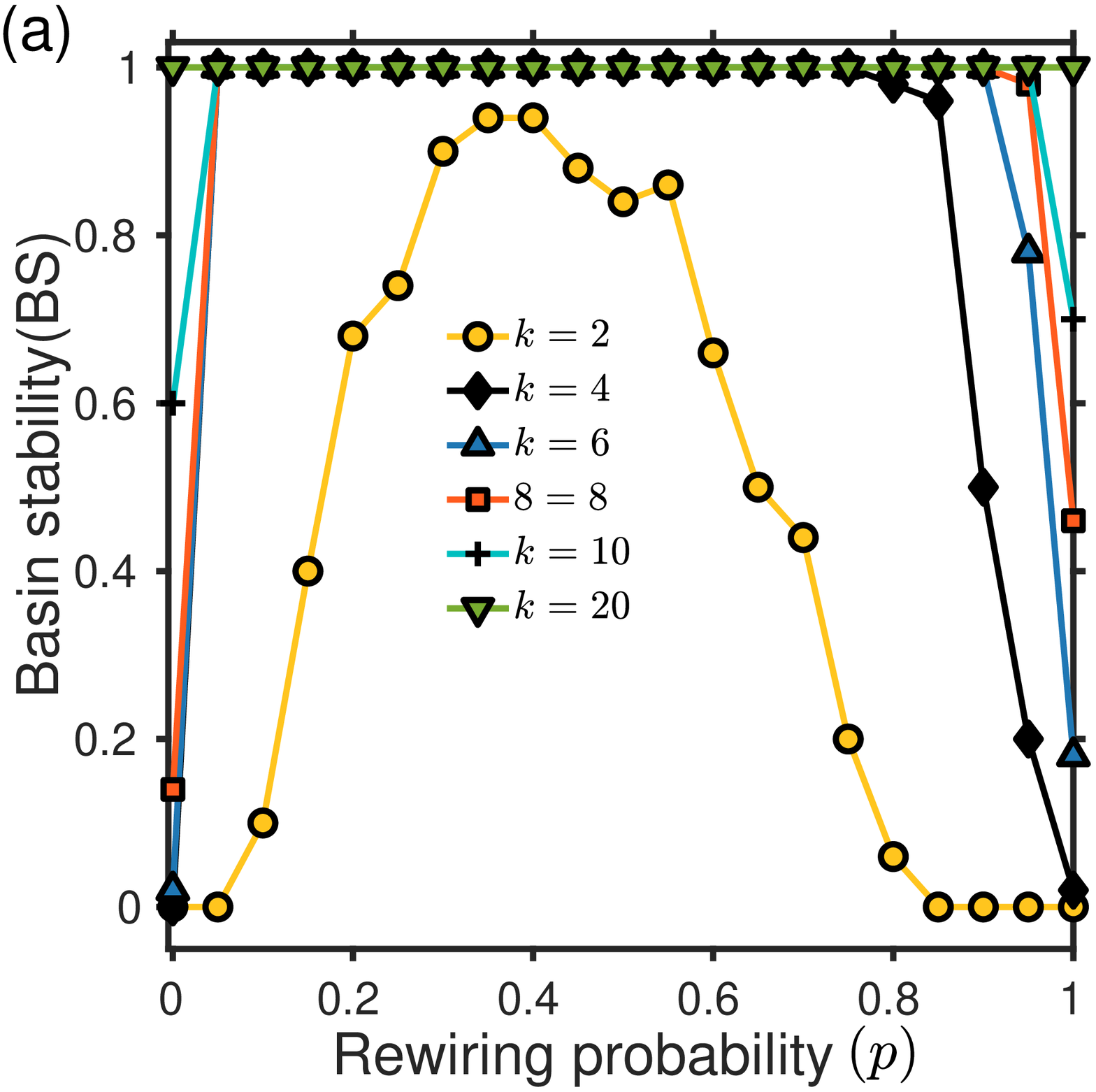}\hspace{0.24in}
\includegraphics[width=0.375\textwidth]{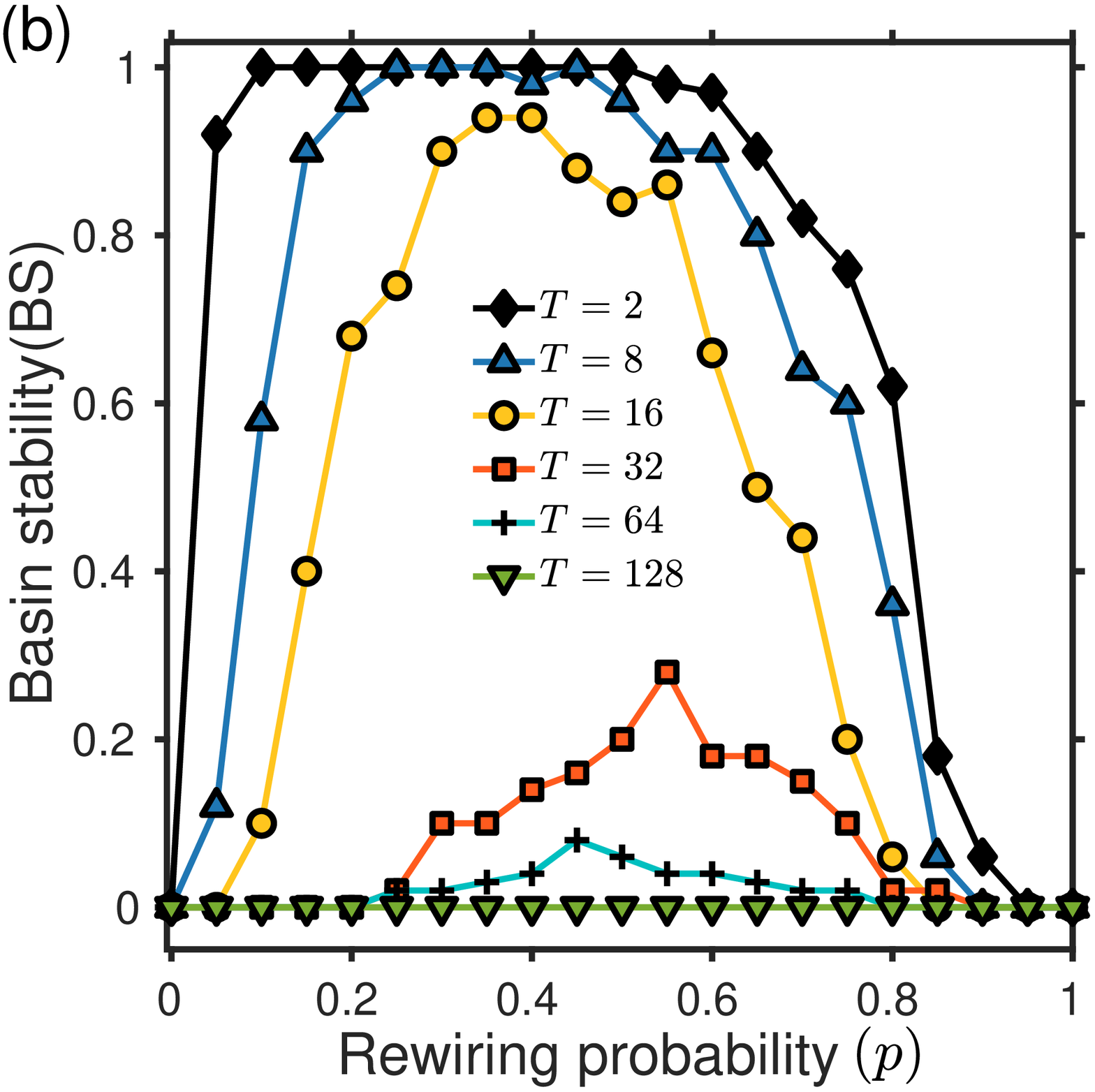}
\caption{Effects of changes in: (a) the average degree $k$ (with $\epsilon=0.005$ and $T=16$), and (b) the rewiring period $T$ (with $\epsilon=0.005$ and $k=2$), on the BS measure of temporal networks. For networks with high average degree ($k = 20$) (i.e., for more connected networks) the BS is almost one irrespective of the chosen rewiring probability $p$.  On increasing $T$ from $T=16$, the BS decreases. However, decreasing $T$ has reverse effects, resulting in higher BS.}
\label{f:smt}
\end{figure*}

Next, we consider a time-varying network structure of the metacommunity with a rewiring period ($T$), where each snapshot of a network follows the WS topology. Here, the rewiring period is considered as $T=16$, which is a sub-harmonics of the Hastings-Powell model's dominant period determined using the wavelet analysis. Here, the BS is computed for varying rewiring probability ($p$) at different values of $\epsilon$. At each value of $p$, a total of $10^3$ simulations is performed with a fixed $\epsilon$ in the time interval $[0,10^4]$. Figure~\ref{f:fig3}(b) shows the BS of the time-varying networks computed for different $p$ values. For a range of $\epsilon$ values, the BS increases on increasing the rewiring probability and then decreases on further increase in $p$ value. In other words, the temporal network with $T=16$ exhibits larger synchronization regions for intermediate values of $p$ and smaller synchronization regions for low and high $p$ values. Hence, in the proximity of regular and completely random network structures, a metacommunity will exhibit higher species persistence by reducing the synchronization region.

Figure~\ref{f:hp_tv_ts} shows the spatial dynamics of the temporal network for different values of $p$. In accordance with the results depicted in Fig.~\ref{f:fig3}(b), depending upon the rewiring probability $p$, here the model displays either asynchronous or synchronous dynamics. For $p=0.01$ and $p=0.99$ the temporal network exhibits asynchrony (see Figs.~\ref{f:hp_tv_ts}(a) and \ref{f:hp_tv_ts}(d) and Figs.~\ref{f:hp_tv_ts}(c) and \ref{f:hp_tv_ts}(f)). However, for $p=0.5$ the synchronous dynamics in the system is easily visible from Figs.~\ref{f:hp_tv_ts}(b) and \ref{f:hp_tv_ts}(e).

\subsection{Effects of the average degree ($k$) and the rewiring period ($T$) on metacommunity persistence}

In this section, we discuss the impact of average degree $k$ and rewiring period $T$ on the collective dynamics of the network. Both of these factors influence the connectivity structure of the metacommunity and hence can significantly influence the population persistence. Figure~\ref{f:fig3}(b) displays that at $\epsilon=0.005$, $k=2$ and $T=16$ the network can exhibit both synchronous and asynchronous dynamics depending upon the rewiring probability $p$. Next, we show that this result significantly depends upon choices of $k$ and $T$. 

To start with, we fix the dispersal rate at $\epsilon=0.005$ and the rewiring period at $T=16$ and determine the BS measure for different values of $k$. With an increase in $k$, the BS increases, resulting in larger synchronization regions (see Fig.~\ref{f:smt}(a)). Eventually, the BS reaches $1$ for a large enough $k$, irrespective of the rewiring probability $p$. Whilst, at higher $k$, the BS does not change depending on $p$, and the network achieves global synchronization regions, at lower $k$ values, the BS exhibits unimodal dynamics along gradients of $p$. Thus, the chance of reaching the synchronous state is more for more connected networks. The BS shown in Fig.~\ref{f:smt}(a) is calculated from $10^3$ independent simulations estimating the frequency of reaching the synchronized state. The result is shown in Fig.~\ref{f:smt}(a) also holds good for different values of the coupling strength $\epsilon$.

Until now, we have considered the rewiring period as $T=16=2^4$, which is a subharmonic of the dominant period $2^7$ as determined by the wavelet analysis (see Fig.~\ref{f:time_phase}). Here, we address how the synchronization region changes with variations in $T$. To calculate the BS measure, we fix $\epsilon=0.005$ and $k=2$.  On increasing $T$ (slower rewiring), from $T=16$, the synchronization regions decrease for different $p$. However, the synchronization regions increase by decreasing $T$ (faster rewiring). This result is depicted in Fig.~\ref{f:smt}(b). Further, we see that when $T=128$, the BS is almost zero irrespective of the rewiring probability $p$. This suggests that if we give the network more time to adapt to the changes in the structure (by increasing the rewiring period $T$), the synchronization regions shrink, resulting in species persistence via asynchrony. 
\begin{figure}[!h]
\centering
\includegraphics[width=0.5\textwidth,angle=0]{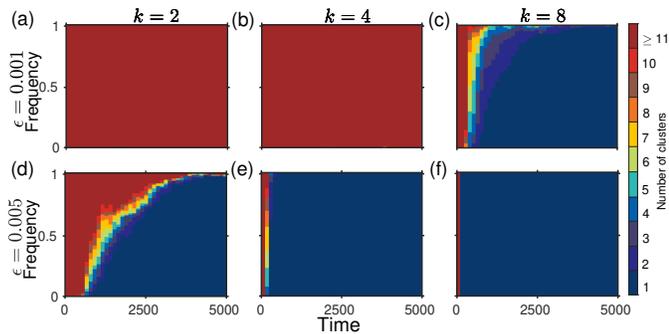}
\caption{Distribution of clustering: ((a)-(c)) At weak ($\epsilon=0.001$), and ((d)-(f)) moderate ($\epsilon=0.005$) dispersal rates with less and more connectivity (left to right panels). The frequency of each cluster is shown using $10^4$ independent simulations. Other parameters are $p=0.2$, and $T=16$.}
\label{f:sm2}
\end{figure}

\subsection{Multi-clustering in time-varying networks}

For a combination of average degree and coupling strength, clusters in the metacommunity \eqref{eq1} are computed with variations in time (Fig.~\ref{f:sm2}). The $N$-patch metacommunity with the time-varying network structure can show $n$-clusters, $1 \leq n \leq N$, which vary over time conditioned by the near neighbor connections ($k$). Here, $N$-cluster represents complete asynchrony (supports species persistence), and $1$-cluster represents complete synchrony (can trigger community collapse and reduce species persistence). Using $10^4$ independent simulations, the frequency of the clusters have been computed and is shown in Figs.~\ref{f:sm2}(a)-\ref{f:sm2}(f) at weak dispersal rate (top panel), moderate dispersal rate (bottom panel), with less number of connections ($k=2$), followed by more number of connections ($k=4$) and ($k=8$). The frequency of $n(\geq 11)$-clusters is high when $k=2$. With increasing $k$, the patches become more synchronous, and we see more $n(\leq 10)$ clusters. Similarly, for a fixed average degree $k$, an increase in the dispersal rate $\epsilon$ increases the synchrony in the systems, and the frequency of 1-cluster solution increases. Hence, a higher dispersal rate and higher average degree are detrimental for metacommunity persistence as they increase the frequency of $n(\leq 10)$-clusters and 1-clusters (global synchrony).

\subsection{Synchronization time of  time-varying networks}

\begin{figure}[!h]
\centering
\includegraphics[width=0.95\columnwidth,angle=0]{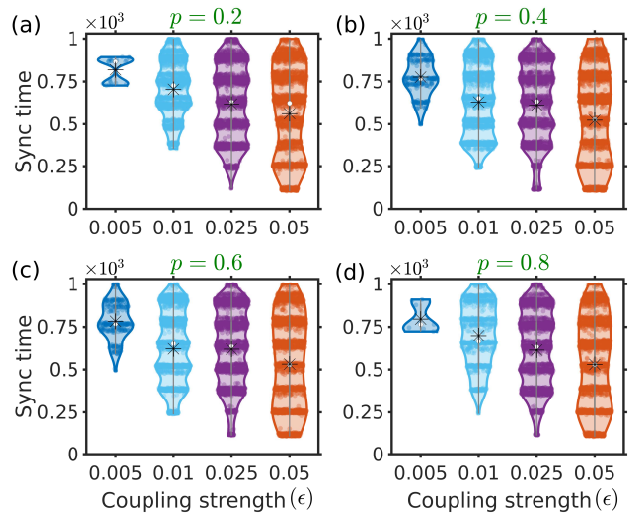}
\caption{Violin plots for synchronization time, corresponding to different $p$ value, for varying coupling strength $\epsilon$. With an increase in $\epsilon$, the number of networks synchronizing with a smaller sync time increases. Also, irrespective of change in $p$, the mean sync time decreases for an increase in $\epsilon$. For each coupling strength, the central black colored mark indicates the mean. A white dot denotes the median, and each violin's bottom and top edges indicate the 25th and 75th percentiles, respectively. The size of a violin represents the initial conditions for which the network synchronizes in the considered time interval. Here, we have considered $5 \times 10^3$ initial conditions to study the sync time for the four different rewiring probabilities ($p=0.2, 0.4, 0.6, 0.8$). Other parameters are $N=100$, $k=2$, and $T=16$.}
\label{f:sm3}
\end{figure}

We calculate synchronization time to assess the influence of network properties in driving the system to synchrony and the time after which it is completely synchronized. The computed synchronization time for networks are represented by violin plots corresponding to each rewiring probability ($p=0.2, 0.4, 0.6$ and $0.8$) (see Fig.~\ref{f:sm3}).  By sorting the synchronization time from least to greatest, we determine the average time in which most networks reach the synchronized state.

In Fig.~\ref{f:sm3}, the minimum and the maximum synchronization time have been indicated by the lower and the upper extremes, respectively. At each rewiring probability, the mean synchronization time of the networks is indicated by the central black mark in each violin. The majority of the networks require less synchronization time with increasing rewiring probability and a further increase in rewiring probability ($p=0.8$), synchronization time increases. The number of synchronized networks increases with rewiring probability and decreases at high rewiring probability. A clear implication from the calculation of synchronization time is at the extremes (very low and very high) rewiring probabilities, the number of networks reaching synchronized state is lesser in agreement with our basin stability measure results in Fig.~\ref{f:fig3}(b). However, results are more prominent at low coupling strength ($\epsilon=0.005$). The results will be qualitatively similar and hold good for different coupling strengths.

\section{Conclusions and Discussion}

The dispersal network structure is an essential factor determining the fate of ecological communities amidst environmental degradation \cite{ranta2007population}. Species may switch interactions and opt for a more viable choice owing to unfavorable habitat conditions in a dynamic environment. Interestingly, these changes can be envisioned in networks at different time scales \cite{zhou2016synchronization}. However, to the best of our knowledge, the dynamics of ecological networks under the framework of a time-varying network topology remains less explored. Extinction in ecological networks has been associated with synchronous dynamics, further increasing risks of a community collapse \cite{earn2000coherence}. Under this backdrop, we study the dynamics of time-varying ecological networks and their impact on metacommunity persistence. Here we take a novel approach of evolving structure in networks for a range of rewiring probabilities with varying rewiring time scales. We obtain an interesting yet alarming result - the time scale of rewiring and the rewiring probability interplay in inducing or dissuading synchrony in the system. Our key results indicate that coupling strength has a positive effect on a certain rewiring probability $p$ leading to synchrony in the system. Post a critical threshold value of $p$, networks tend to be more random, and the system reaches an asynchronous state. One of the main results of our study is that the slower rewiring periods promote asynchrony in the system. We observe that on increasing the rewiring period, the BS decreases irrespective of the rewiring probability and eventually pushes the system to an asynchronous state. Apart from the basin stability measure, the estimated synchrony time and multi-frequency cluster analysis support our key findings.

Our work presents an in-depth study of collective population dynamics in temporal networks using the MSF approach and the BS measure that aids in investigating local and global synchrony, respectively. Certainly, quantifying the stability of the synchronous manifold is of grave ecological importance. While in the face of global environmental change, the evolution of species dispersal network structure is inevitable, our results indicate that slowing the evolutionary time-scale can serve as a mitigation strategy to prevent synchrony -- thus reducing global extinction risk. We believe our results have much broader implications for managing real ecological networks and demand further in-depth research in this direction. We validate the robustness of our results for another important ecological model, namely, the Blasius-Huppert-Stone model \cite{blasius1999complex} (see Appendix). We obtain qualitatively similar findings for both models. Our approach provides intriguing results, albeit requiring future investigation in a large class of other ecological networks. While structural evolution is obligatory across networks of diverse origin, such as biogeochemical networks \cite{falkowski1998biogeochemical}, food-trade networks \cite{wang2021evolution}, and other socio-economic networks \cite{schweitzer2009economic,liu2012structure}, further work along this direction can provide practical mitigation policies towards a sustainable future.

\section*{Acknowledgments}
P.S.D. acknowledges financial support from SERB, Department of Science and Technology (DST), India (Grant number: CRG/2019/002402). S.B. acknowledges Ramesh Arumugam for helpful discussion.

\section*{APPENDIX: The Blasius-Huppert-Stone metacommunity model}

\begin{figure}[htpb]
\centering
\includegraphics[width=0.9\columnwidth,angle=0]{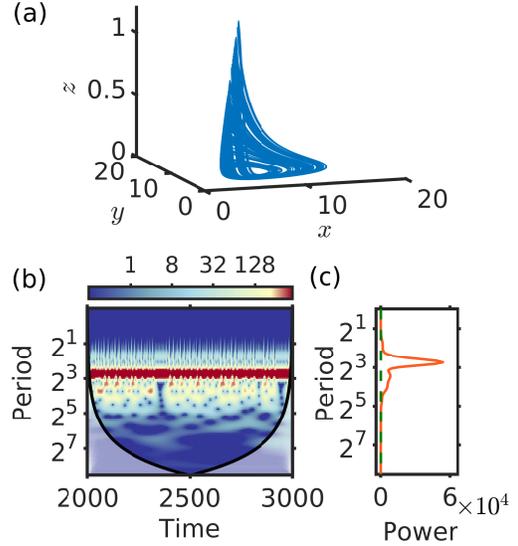}
\caption{Wavelet analysis to a chaotic time series of the metacommunity model \eqref{eqA1}: (a) Phase-portrait depicting a chaotic trajectory, (b) corresponding wavelet power spectra, and (c) the wavelet global spectrum. Model parameters are $a=1$, $b=1$, $c=10$, $\beta_1=0.2$, $\beta_2=1$, $K_{1}=0.05$, and $w^*=0.006$.} \label{f:sm4}
\end{figure}

\begin{figure*}
\centering
\includegraphics[width=0.85\textwidth,angle=0]{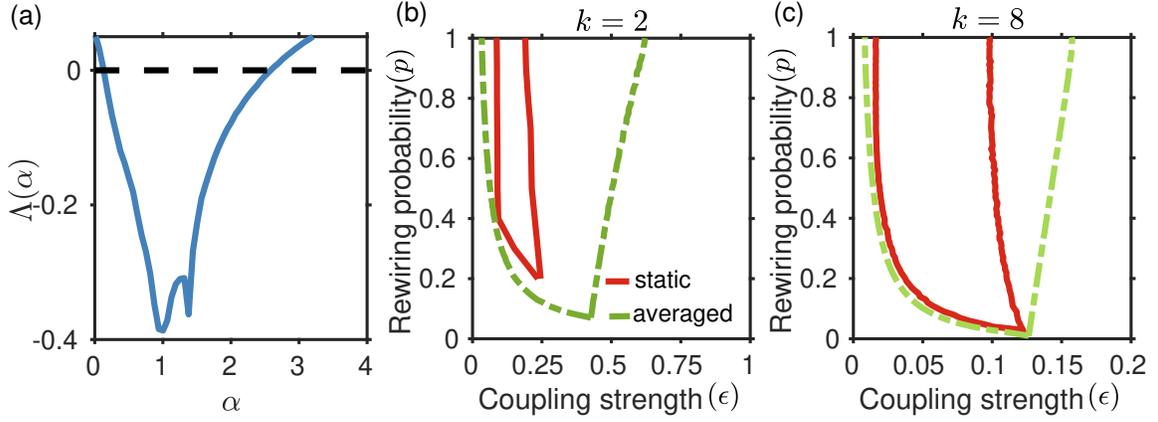}
\caption{(a) Master stability function for the coupled Blasius-Huppert-Stone model \eqref{eqA1}. The black dashed line marks the neutral line.  The region of stable synchronized solution is plotted as a function of $p$ and $\epsilon$ in solid (dashed) curve corresponding to the static (averaged) network calculated using the MSF approach; for (b) $k = 2$ and (c) $k = 8$. The region between solid (dashed) curves corresponds to the range of coupling strength $\epsilon$ where the synchronous state is stable for the static (averaged) network.}
\label{f:smt3}
\end{figure*}

\begin{figure*}
\centering
\includegraphics[width=0.38\textwidth,angle=0]{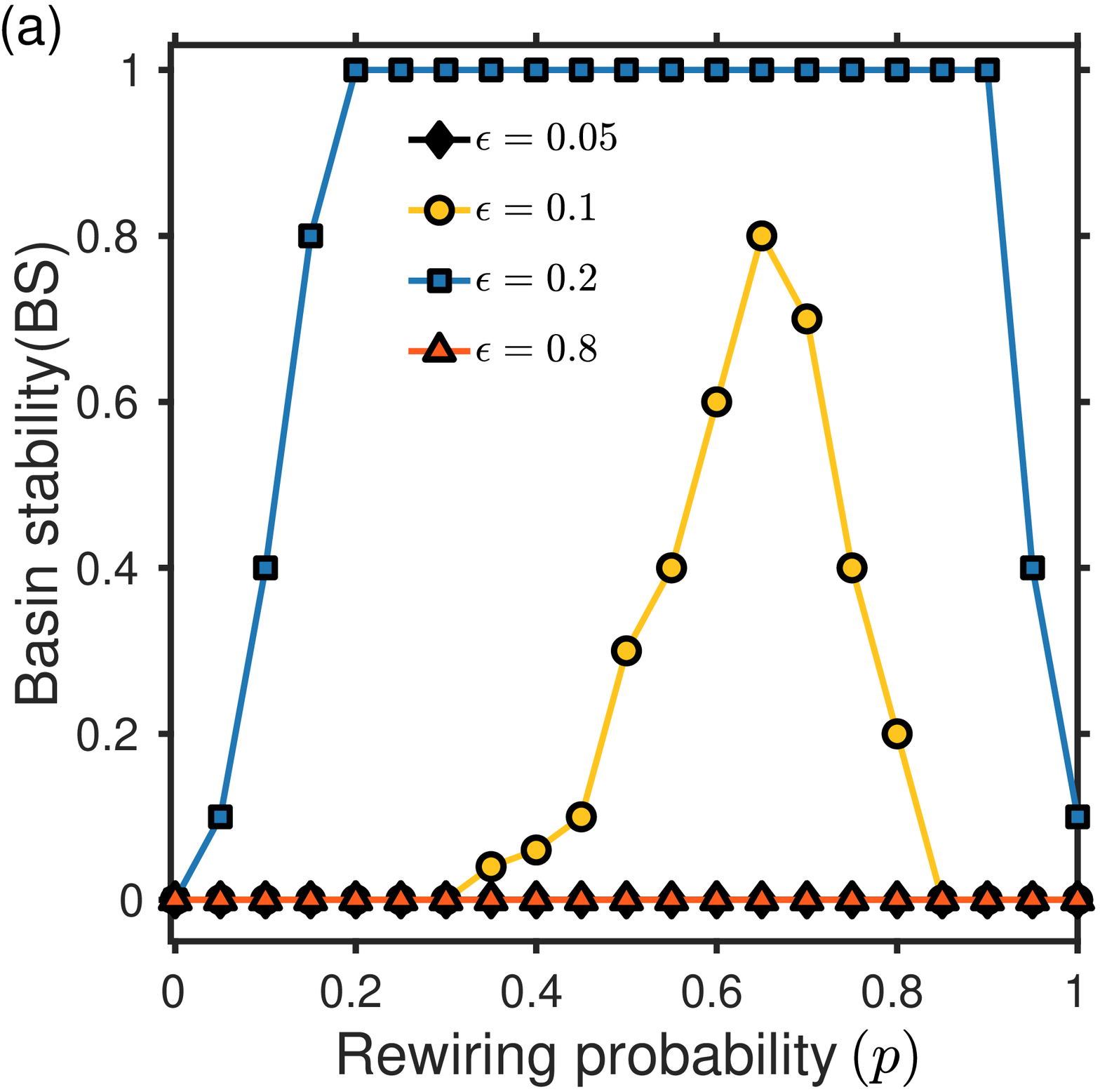} \hspace{0.2in}
\includegraphics[width=0.3748\textwidth,angle=0]{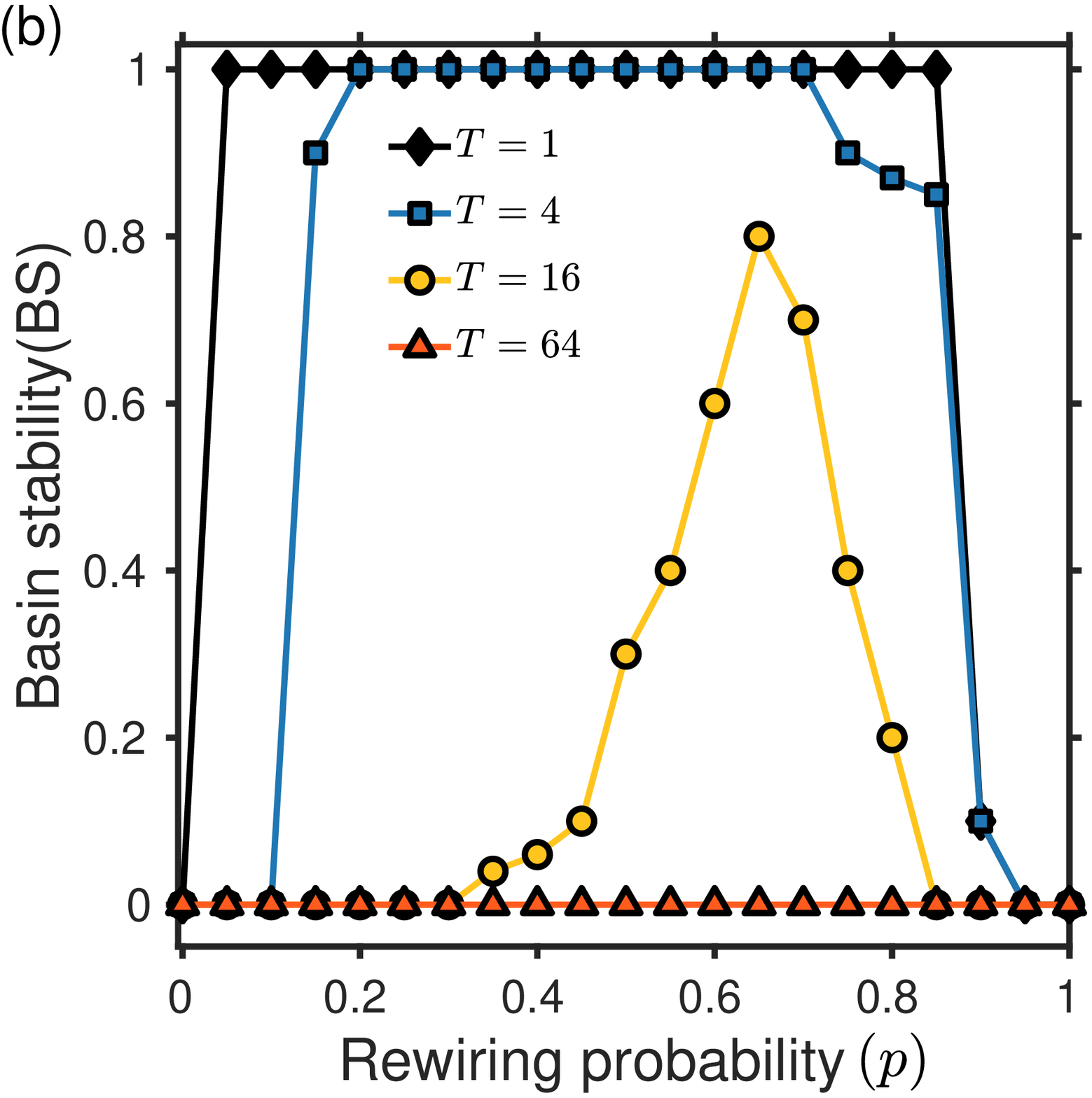}
\caption{Basin stability (BS) of the time-varying network \eqref{eqA1} across different values of the rewiring probability $p$: (a) For different values of $\epsilon$ (with $k=2$ and $T=16$), and (b) for different values of $T$ (with $k=2$ and $\epsilon=0.1$).} \label{f:BSAp}
\end{figure*}

We demonstrate results for the MSF approach and the BS regions for another temporal ecological network model - the Blasius-Huppert-Stone model \cite{blasius1999complex}. The results obtained are qualitatively similar to the Hastings-Powell model and add to the generality of our study. The coupled Blasius-Huppert-Stone network model is represented as follows:
\begin{subequations}\label{eqA1}
\begin{align} 
\frac{dx_i}{dt}&= a x_i-\beta_{1}\frac{x_i y_i}{1+K_1 x_i},\\ 
\frac{dy_i}{dt}&= \beta_{1}\frac{x_i y_i}{1+K_1 x_i}-\beta_{2} y_i z_i-b y_i +\epsilon \sum_{j=1}^{N} L_{ij}y_j,\\ 
\frac{dz_i}{dt}&= -c(z_{i}-w^{*})+ \beta_{2} y_i z_i +
\epsilon \sum_{j=1}^{N} L_{ij} z_j,
\end{align} 
\end{subequations}
where $x_{i}$, $y_{i}$, $z_{i}$ represents vegetation, herbivore and predator populations, respectively, in the $i$-th patch. The growth rates of each trophic species in the absence of interspecific interaction are represented by the parameters $a$, $b$ and $c$, respectively. The predator-prey and consumer-resource interactions are incorporated into the equation via the Lotka-Volterra term or the Holling type-II interaction term. $\epsilon$ denotes  the dispersal rate. When $\epsilon=0$, for a specific set of parameters dynamics of the model \eqref{eqA1} are chaotic, and the attractor is displayed in Fig.~\ref{f:sm4}(a). Corresponding wavelet analyses, to determine the rewiring period $T$, are presented in Figs.~\ref{f:sm4}(b)-\ref{f:sm4}(c). Parameters of the uncoupled model \eqref{eqA1} (when $\epsilon=0$) used for numerical simulations are $a=1$, $b=1$, $c=10$, $\beta_1=0.2$, $\beta_2=1$, $K_{1}=0.05$, and $w^*=0.006$.

We have calculated the stability regions of synchronous state using the MSF approach for static and temporal networks as shown in  Fig.~\ref{f:smt3}. Figure~\ref{f:smt3}(a) shows expected stability intervals for varying normalized coupling strength ($\alpha$). The range of the coupling strength ($\epsilon$) in which a synchronous solution is stable for different rewiring probability and average degree ($k=2$ and $k=8$) are plotted in Figs.~\ref{f:smt3}(b)-\ref{f:smt3}(c). We observe that the synchronous state is stable in $\alpha_{1}<\epsilon \lambda_{k}< \alpha_{2}$, where $\alpha_{1}=0.13$ and $\alpha_{2}=2.62$. One can also conclude that the expected regions of stable synchronous solution decrease with decreasing rewiring probability $p$, and the result is similar to the one shown in Fig.~\ref{f:smt1}. While, in Fig.~\ref{f:smt1} the stability region is bounded below only, here in Fig.~\ref{f:smt3} it is bounded both below and above.

Figure~\ref{f:BSAp} exhibits changes in the BS for different coupling strengths $\epsilon$ and rewiring period $T$, with $k=2$. In Fig.~\ref{f:BSAp}(a), we observe that for moderate $\epsilon$ values, the synchronous solution is stable for intermediate rewiring probabilities. However, for low and high $\epsilon$ values, the BS is zero irrespective of the choice of rewiring probability $p$, resulting in complete asynchrony in the system. This result is in agreement with the synchronization region calculated using the MSF approach (see Fig.~\ref{f:smt3}(b)). Further, increasing $T$ lowers the BS of the time-varying networks (Fig.~\ref{f:BSAp}(b)), and eventually, the BS becomes zero for all values of $p$ at a high rewiring period $T$. These results are in line with our previous findings illustrated in Fig.~\ref{f:fig3}(b) and Fig.~\ref{f:smt}(b) for the Hastings-Powell model. 


%

\end{document}